\documentclass[preprint,authoryear, 12pt]{elsarticle}
\usepackage{setspace}
\usepackage{color}
\usepackage{lipsum}
\usepackage{amsfonts,amsmath,amssymb,mathrsfs}
\usepackage{epstopdf}
\usepackage{natbib}
\usepackage{algorithmic}
\usepackage{amsopn}
\usepackage{mathtools}
\usepackage{cases}
\usepackage{color}
\usepackage{enumerate}
\usepackage{graphicx}
\usepackage{subcaption}
\newtheorem{theorem}{Theorem}[section]
\newtheorem{lemma}{Lemma}[section]
\newtheorem{definition}{Definition}[section]
\newtheorem{remark}{Remark}[section]
\newtheorem{assumption}{Assumption}

\newtheorem{proposition}{Proposition}[section]
\newtheorem{corollary}{Corollary}[section]

\usepackage{geometry}
\renewcommand{\epsilon}{\varepsilon}

\numberwithin {equation} {section}

\begin{document}
	\journal{arXiv.org}
\begin{frontmatter}
\title{Robust distortion risk measures with linear penalty \\ under distribution uncertainty}

\address[math]{School of Mathematics, China University of Mining and Technology, P.R. China}

\author[math]{Yuxin Du}
\ead{18203431198@163.com}

\author[math]{Dejian Tian\corref{correspondingauthor}}
\ead{djtian@cumt.edu.cn}
\cortext[correspondingauthor]{Corresponding author}

\author[math]{Hui Zhang}
\ead{zhanghui\_cumt@126.com}

\begin{abstract}
The paper investigates the robust distortion risk measure with linear penalty function under distribution uncertainty.  The distribution uncertainties are characterized by predetermined moment conditions or constraints on the Wasserstein distance.  The optimal quantile distribution and the optimal value function are explicitly characterized.  Our results partially extend the results of \cite{bpv24} and \cite{l18} to  robust distortion risk measures with linear penalty. In addition, we also discuss the influence of the penalty parameter on the optimal solution. 
\end{abstract}

\end{frontmatter}

\section{Introduction}


Traditional risk measures, such as variance, are insufficient to address extreme risks. To tackle this, distortion risk measures have been developed. By ``distorting" the risk distribution and emphasizing tail risks, this approach enables more accurate assessments of potential extreme losses, especially in volatile markets.  Distortion risk measures, particularly Value-at-Risk (VaR) and Conditional VaR (CVaR), are widely used in portfolio optimization and risk management. The reader can refer to seminal documents such as \cite{ya87}, \cite{WYP97}, \cite{w96} and \cite{ad99}, and academic textbooks \cite{fs16}. 

Meanwhile, in the financial domain, investment decisions and risk management face significant distribution uncertainty due to market fluctuations, economic changes, and external shocks, particularly in parameters such as asset returns, interest rates, and exchange rates. The theory of risk measures offers a comprehensive framework for managing uncertainties and extreme risks in financial markets when integrated with robust optimization problems for portfolio optimization, risk management, and asset pricing. 
The relevant literature includes \cite{ek18}, \cite{gk23}, \cite{gx14},  and \cite{bd20}. 

Owing to their distinctive mathematical properties, distortion risk measures have also been increasingly utilized in the study of distributionally robust optimization frameworks, particularly when addressing uncertainties in probability distributions. \cite{l18} investigates law invariant risk measures that evaluate maximum risk is based on limited information about the underlying distribution,  and provides closed-form solutions for worst-case law invariant risk measures.  Furthermore, \cite{cl23} and \cite{pw24} explore worst-case scenarios under distortion risk measures, where the closedness under concentration or convex hull techniques is used for the  non-convex problems.

 \cite{bpv24} focus on optimization problems for robust distortion risk measures under distribution uncertainty by analyzing their robustness amid parameter uncertainty and volatility. They quantify the robustness of distortion risk measures with absolutely continuous distortion functions under distributional uncertainty by evaluating the maximum (or minimum) value of the loss distribution, characterized by its mean and variance, within a domain defined around the reference distribution using the Wasserstein distance. In addition,
\cite{hcm24} consider the robust optimization problem for the expectile risk measures, while \cite{pj23} and \cite{bc22} investigate robust optimization problems for the mean-variance model and active portfolio management using the Wasserstein distance.

This study introduces a \textit{penalty term} for the distance between the target and reference distributions within the framework of distortion risk measures.  Another motivation is from the comonotonic convex risk measures proposed by \cite{sy09}, the present study primarily investigates the following optimization problem:
\begin{equation}\label{eq:sec1.1}
	\mathop{\sup}_{G\in \mathcal{N}}\ H_g (G)-{\varphi({d^2_W(F,G)})},
\end{equation}where $g$ is a distortion function, $\varphi$ is a predetermined linear penalty function and $\mathcal{N}$ can be chosen as distribution uncertainty sets $\mathcal{M}(\mu,\sigma)$ with given moment conditions or  $\mathcal{M}_{\varepsilon}(\mu,\sigma)$ with constraints on Wasserstein distance (see more details in section \ref{sec:2}).
By incorporating a distance-based penalty, this approach increases the emphasis on extreme losses, encouraging decision-makers to prioritize severe risks over average or likely losses.

The paper contributes to the literature in the following four aspects.  First, the optimal quantile distribution and its corresponding optimal value are explicitly obtained for $\mathcal{M}(\mu,\sigma)$ with concave distortion functions. Theorem \ref{t1}  extends the results of \cite{l18} by admitting the linear penalty function.  In particular,  when the penalty coefficient is zero, our result aligns with the findings in \cite{l18}. 
In addition, Proposition \ref{pro:emax} provides a sharper estimate between the optimal quantile distribution and the reference distribution, which is very helpful to consider the optimization problem with $\mathcal{M}_{\varepsilon}(\mu,\sigma)$.

Second, we completely solve the optimal quantile function and its corresponding optimal value under $\mathcal{M}_{\varepsilon}(\mu,\sigma)$ for different combinations of penalty coefficient and distance parameter with concave distortion functions.  Given that distance and penalty coefficients are interrelated parameters, we investigate their functional relationship, discovering a one-to-one correspondence between them, which enables us to establish different boundaries for determining the optimal quantile and its optimal value in various cases. 

Theorem \ref{tu-2} indicates that the agent makes a \textit{trade-off} between the distortion risk measure and the penalty term, depending on the Wasserstein distance and the penalty parameter. Especially, when the penalty parameter is less than the critical value, the optimal quantile distribution is chosen on the boundary of the Wasserstein ball, not in the interior any more.  This characteristic is the key ingredient between our model and the model of \cite{bpv24}.  When the penalty coefficient is zero, the results are consistent with the conclusions in \cite{bpv24}. Some graphic analysis is also carried out for discussion about the penalty parameter and comparison with the results of \cite{bpv24} and \cite{l18}.

Third, we extend our analysis to the general distortion function cases.  Motivated by the isotonic projection technique employed by \cite{bpv24}, under some milder conditions, Theorem \ref{t41} and Theorem \ref{th-hat} successfully obtain the optimal quantile distributions in distribution uncertainty sets $\mathcal{M}(\mu,\sigma)$ and $\mathcal{M}_{\varepsilon}(\mu,\sigma)$ for the  general distortion function, respectively. 

Finally, we apply our theoretical findings to the practical context of CVaR. We derive explicit solutions and visualize the results, illustrating the impacts of varying distances and penalty coefficients on risk measures. These visualizations provide a more precise and intuitive understanding, facilitating a better grasp of how distance and penalty coefficients influence the outcomes of risk measures.

The structure of this paper is as follows. Section \ref{sec:2} introduces distortion risk measures and the formulation of optimization problem. Section \ref{sec:3} solves the optimization problem for the distortion risk measure with penalty under distribution uncertainty in the situation of concave distortion functions.
The non-concave case is discussed in Section \ref{sec:4}. 
Section \ref{sec:5} applies the results to the model of CVaR, and Section \ref{sec:6} concludes the paper.

\section{Model setup and problem formulation}\label{sec:2}

Let $(\Omega, \mathcal{F}, P)$ be an atomless probability space. Let $L^2=L^2(\Omega, \mathcal{F}, P)$ be the set of all square-integrable random variables.  
Denote $\mathcal{M}^2=\left\{F(\cdot)=P(X\le \cdot~)~|~X\in L^2\right\} $ for the distribution functions with finite second moment.  In particular, we write $U\sim U(0,1)$
for a standard uniform random variable on $(0,1)$. 
For any $F\in \mathcal{M}^2$, define its left-continuous inverse (or quantile function) as follows: 
$$F^{-1}(u)=\inf{\left\{ y\in \mathbb{R}~|~F(y)\ge u\right\}}, ~~\forall u\in(0,1).$$By convention, $\inf\emptyset=+\infty$.  


\subsection{Distribution uncertainty}
For any $G_1,G_2\in\mathcal{M}^2$, 
recall the second order Wasserstein distance (see \cite{v09}):  
$$d_W(G_1,G_2)=\inf\big\{(\mathbb{E}[(X-Y)^2])^{1/2}~\vert~ X\sim G_1, ~Y\sim G_2\big\}=\left(\int_0^1(G_1^{-1}(u)-G_2^{-1}(u))^2\text{d}u\right)^{1/2}.$$
We use $d_W(\cdot,\cdot)$ to characterize the distance or  discrepancy of two distributions, and it is determined by their corresponding
quantile functions.

For any $\varepsilon>0$, $\mu\in\mathbb{R}$, $\sigma>0$, we consider the following distribution uncertainty sets:
\begin{align*}
    \mathcal{M}(\mu,\sigma)&=\left\{G\in{\mathcal{M}}^2 ~\big|~ \int x~\text{d}G(x)=\mu,\int x^2~\text{d}G(x)={\mu}^2+{\sigma}^2\right\};\\
    \mathcal{M}_\varepsilon (\mu ,\sigma) &=  \left\{G\in{\mathcal{M}}^2 ~\big|~\int x~\text{d}G(x)=\mu,\int x^2 ~\text{d}G(x)={\mu}^2+{\sigma}^2 ,d_W(F,G)\le\sqrt{\varepsilon}\right\}.
\end{align*}
The set $\mathcal{M} (\mu, \sigma) $ contains all distribution functions whose first two moments are $\mu$ and ${\mu}^2+{\sigma}^2$ respectively.  The set $\mathcal{M}_\varepsilon (\mu,\sigma)$ contains all distribution functions whose first two moments are $\mu$ and ${\mu}^2+{\sigma}^2$, and the distribution function $G$ is located in a Wasserstein sphere less than $\sqrt{\varepsilon}$ from the reference distribution $F$.

For the given $F\in{ \mathcal{M}}^2 $ as a reference distribution, suppose that we know its first two moments as follows:  
$$\int x\text{d}F(x)=\mu_F\in \mathbb{R}, ~~\int x^2 \text{d}F(x)={\mu}_F^2+{\sigma}_F^2,~{\sigma}_F>0.$$We can get a more explicitly expression for the Wasserstein distance,  for any  $G\in \mathcal{M}(\mu,\sigma)$, 
\begin{align}\label{eq:dfg}
    {d_W^2(F,G)}&=\int_0^1 {(F^{-1}(u)-G^{-1}(u))}^2\text{d}u\nonumber\\
		&=(\mu_F-\mu)^2+(\sigma_F-\sigma)^2+2\sigma\sigma_F(1-\text{corr}(F^{-1}(U),G^{-1}(U))).
\end{align}
Noting that the distance between the reference distribution function $F$ and the desired distribution function $G$ is bounded, to be specifically,  
\begin{align}\label{eq:emin-emax}
	\varepsilon_{min}&\le {d_W^2(F,G)}\leq \varepsilon_{min}+2\sigma\sigma_F, ~~~\forall G\in\mathcal{M}(\mu,\sigma),
\end{align}where $\varepsilon_{min}:=(\mu_F-\mu)^2+(\sigma_F-\sigma)^2$.  
Therefore, $\mathcal{M}_{\varepsilon}(\mu,\sigma)=\emptyset$ if $\varepsilon_{min}>\varepsilon$.

\subsection{Problem formulation}
The function $g:[0,1]\rightarrow[0,1]$ is called a distortion function if it is non-decreasing and satisfies $g(0)=0$ and $g(1)=1$. For some given distortion function $g$, define a distortion risk measure by the following Choquet integral: 
\begin{equation*}%
		H_g (G)=\int_0^{\infty}g(1-G(x)){\rm d}x+\int_{-\infty}^0\big(g(1-G(x))-1\big){\rm d}x,~~\forall G\in \mathcal{M}^2,
\end{equation*}whenever at least one of the two integrals is finite. When $g$ is absolutely continuous, then $H_g (\cdot)$ can also be written as a spectral risk measure (\cite{d12}, Theorem 6): 
\begin{equation}\label{f1}
H_g (G)=\int_0^1\gamma(u)G^{-1}(u)\text{d}u,~~\forall G\in\mathcal{M}^2, 
\end{equation}
where $\gamma(u)=\partial^-g(x)|_{x=1-u},0<u<1$, which satisfying $\int_0^1\gamma(u)\text{d}u=1$. 

To guarantee the finiteness or non-trivial of \eqref{f1}, we impose the following assumption on the distortion function $g$.

\begin{assumption}\label{a1}
Suppose equation \eqref{f1} holds,
and $$\int_0^1\gamma^2(u)\text{d}u<+\infty~~\text{and}~~ \sigma_0:=\textrm{std}(\gamma(U))>0,$$ where $U\sim U(0,1)$. 
\end{assumption}

Based on the distortion risk measure, we also require that the distance between the desired distribution and the reference distribution cannot be too far. In order to control and emphasize the distance between the desired distribution and the reference distribution, we impose a penalty on their distance based on the distortion risk measure. Our problem is to find the optimal distribution for the distortion risk measure with some penalty function $\varphi$ under the uncertainty sets of the distribution.

More precisely, for some given $F\in\mathcal{M}^2$ and distortion function $g$,  our problems are as follows:
\begin{equation}\label{m1}
	\mathop{\sup}_{G\in \mathcal{N}}\ H_g (G)-{\varphi({d^2_W(F,G)})},
\end{equation}where $\varphi:[0,\infty)\rightarrow[0,+\infty]$ is a predetermined penalty function and $\mathcal{N}$ can be chosen as $\mathcal{M}(\mu,\sigma)$ and  $\mathcal{M}_{\varepsilon}(\mu,\sigma)$ respectively.  In addition, we want to find the distribution function that achieves the best case. 

In order to make our problem more solvable, this paper
focuses on the linear penalty function. The other types of penalty functions are left for future study.  

\begin{assumption}\label{a2}
The penalty function $\varphi(x)=\delta x$ , $x\geq0$, where penalty parameter $\delta\geq0$.
\end{assumption}

\begin{remark}
    When there is no penalty function (i.e., $\varphi=0$ or  $\delta=0$),  \cite{bpv24} recently considered the situation of $\mathcal{N}=\mathcal{M}_{\varepsilon}(\mu,\sigma)$ and \cite{l18} investigated the case $\mathcal{N}=\mathcal{M}(\mu,\sigma)$. 
\end{remark}

\begin{remark}
From the point of view of the risk measure, another motivation for the study of the distortion risk measure with penalty function is the comonotonic convex risk measures proposed by \cite{sy09}. This kind of risk measures are also studied by \cite{x13}, \cite{TJ15} and \cite{hwwx21}.  
\end{remark}

We end this section with some  notations, which are used later. 
\begin{align}\label{eq: notations}
    \rho:=\text{corr}(F^{-1}(U),\gamma(U)), 
 \qquad\varepsilon_{max}=\varepsilon_{min}+2\sigma\sigma_F(1-\rho).   
\end{align}

\section{Main results}\label{sec:3}

This section solves the  optimization problem \eqref{m1} with the concave distortion function. The situation of non-concave distortion function will be discussed in Section \ref{sec:4}. Subsection \ref{subsec:3.1} gives the solution for $\mathcal{N}=\mathcal{M}(\mu,\sigma)$, and subsection \ref{subsec:3.2} presents the solution for $\mathcal{N}=\mathcal{M}_{\varepsilon}(\mu,\sigma)$ and depicts some figures for discussions.  

\subsection{ $\mathcal{N}=\mathcal{M}(\mu,\sigma)$}\label{subsec:3.1}


The following theorem provides the optimal distribution result of problem \eqref{m1} for $\mathcal{N}=\mathcal{M}(\mu,\sigma)$. The result extends Theorem 2 in \cite{l18} to admit the linear penalty function.



\begin{theorem}\label{t1}
Let $g$ be a concave distortion function and $\mathcal{N}=\mathcal{M}(\mu,\sigma)$. Suppose \ref{a1} and \ref{a2} hold.   Then problem \eqref{m1} has a unique solution, and its optimal quantile function is
\begin{equation}\label{eq:optimal-quantile}
	G^{-1}(u)=\mu-\frac{\sigma}{\sigma_\delta}\mu_\delta+\frac{\sigma}{\sigma_\delta}(\gamma(u)+2{\delta }F^{-1} (u)),~~0<u<1,
\end{equation}
where $\mu_\delta=\mathbb{E}[\gamma(U)+2\delta F^{-1} (U)]$, $\sigma _\delta=\text{std}(\gamma(U)+2\delta F^{-1} (U))$. Furthermore, the corresponding optimal value function is
\begin{align*}
H_g (G)-\varphi({d_W^2(F,G)} )=\mu-\delta(\varepsilon_{min}+2\sigma\sigma_F)+\sigma\sigma_{\delta}.
\end{align*}
\end{theorem}

\noindent\textbf{Proof}.  For any $G\in\mathcal{M}(\mu,\sigma)$, recalling $U\sim U(0,1)$,  one can obtain that
\begin{align}\label{eq:erg}
H_g(G)=\mathbb{E}[\gamma(U)\cdot G^{-1}(U)]&=\mathbb{E}[\gamma(U)]\mathbb{E}[G^{-1}(U)]+\text{cov}(\gamma(U),G^{-1}(U))\nonumber\\
&=\mu+\sigma\sigma_0\text{corr}(\gamma(U),G^{-1}(U)),  
\end{align}where the third equality is derived from the facts that 
$\mathbb{E}[\gamma(U)]=1$, $\mathbb{E}[G^{-1}(U)]=\mu$ and $\text{std}(G^{-1}(U))=\sigma$, respectively.  

Using equations \eqref{eq:dfg}, \eqref{f1} and  \eqref{eq:erg}, by  directly calculations, then the objective function of problem \eqref{m1} can be reduced to\begin{align*}
&H_g (G)-{\varphi({d_W^2(F,G)} )}\\
=&\mathbb{E}[\gamma(U)\cdot G^{-1}(U)]-\delta(\varepsilon_{min}+2\sigma\sigma_F(1-\text{corr}(F^{-1}(U),G^{-1}(U)))\\
=&\mu+\text{cov}(\gamma(U), G^{-1}(U))-\delta(\varepsilon_{min}+2\sigma\sigma_F)+2\delta\text{cov}( F^{-1}(U),G^{-1}(U))\\
=&\mu-\delta(\varepsilon_{min}+2\sigma\sigma_F)+\sigma \text{std}(\gamma(U)+2\delta F^{-1}(U))\text{corr}(\gamma(U)+2\delta F^{-1}(U),G^{-1}(U)).
\end{align*}
From the above, one recognizes that if $\text{corr}(\gamma(U)+2\delta F^{-1}(U),G^{-1}(U))=1$, then the objective function can obtain the maximum value.  In other words, when $\gamma(U)+2\delta F^{-1}(U)$ is completely positive correlated with $G^{-1}(U)$, taking account of the first two moment constraints, then the optimal value can be obtained by choosing the distribution $G$ with quantile function 
\begin{equation*}
	G^{-1}(u)=\mu-\dfrac{\sigma}{\sigma_\delta}\mu_\delta+\dfrac{\sigma}{\sigma_\delta}(\gamma(u)+2{\delta }F^{-1} (u)),~~0<u<1,
\end{equation*}where $\mu_\delta=\mathbb{E}[\gamma(U)+2\delta F^{-1} (U)]$, $\sigma _\delta=\text{std}(\gamma(U)+2\delta F^{-1} (U))$.

Therefore, with the expression above for $G^{-1}$, the corresponding optimal value is$$H_g (G)-\varphi({d_W^2(F,G)} )=\mu-\delta(\varepsilon_{min}+2\sigma\sigma_F)+\sigma \sigma_{\delta}.$$The proof is complete.  \hfill$\Box$

\begin{remark}
Under Assumption \ref{a1}, it is obviously $\sigma_{\delta}=\text{std}(\gamma(U)+2\delta F^{-1}(U))>0$.  
Despite the existence of penalty functions, the proof procedure of Theorem \ref{t1} is standard.
The form of the optimal quantile function \eqref{eq:optimal-quantile} is very important for the subsequent proof.   
\end{remark}

Theorem \ref{t1} indicates that the penalty parameter $\delta$ significantly affects the value function, while structure of the optimal solution does not change much. In particular, when there is no penalty function ($\delta=0$), the result of Theorem \ref{t1} reduces to the Theorem 2 in \cite{l18}.

\begin{corollary}\label{c1}
(\cite{l18})
Suppose the Assumptions in Theorem \ref{t1} hold. When $\delta=0$, then  
the optimal quantile is $$G^{-1}(u)=\mu+\frac{\sigma}{\sigma_0}(\gamma(u)-1),~~u\in(0,1),$$ where $\sigma_0=\text{std}(\gamma(U))$, and the optimal value function is
\begin{align*}
   \mathop{\sup}_{G\in \mathcal{M}(\mu,\sigma)} H_g (G)=\mu+\sigma \sigma_0.
\end{align*}
\end{corollary}

Theorem \ref{t1} gives the quantile formulation of the optimal distribution among in $\mathcal{M}(\mu,\sigma)$.  The next proposition investigates the boundedness  of the distance between this optimal distribution and the reference distribution, which is more sharper than \eqref{eq:emin-emax} under the situation of concave distortion function. Besides, this distance is very useful when we consider the problem \eqref{m1} with $\mathcal{N}=\mathcal{M}_{\varepsilon}(\mu,\sigma)$.

\begin{proposition}\label{pro:emax}
Let $g$ be a concave distortion function and suppose \ref{a1} and \ref{a2} hold. 
For any $\delta>0$, suppose the quantile function for distribution function $G_\delta$ defined as follows:
\begin{equation}\label{eq:optimal-quantile-g-delta}
	G^{-1}_{\delta}(u):=\mu-\dfrac{\sigma}{\sigma_\delta}\mu_\delta+\dfrac{\sigma}{\sigma_\delta}(\gamma(u)+2{\delta }F^{-1} (u)),~~0<u<1,
\end{equation}
where $\mu_\delta=\mathbb{E}[\gamma(U)+2\delta F^{-1} (U)]$, $\sigma _\delta=\text{std}(\gamma(U)+2\delta F^{-1} (U))$.  
Then $G_\delta \in\mathcal{M}(\mu,\sigma)$, and the Wasserstein distance between $G_\delta$ and the reference distribution $F$ satisfies
\begin{align}\label{eq:optimal-distance}
\varepsilon_{min}\leq d_W^2(F,G_{\delta})\leq \varepsilon_{max}, ~~\forall \delta>0,
\end{align}where $\varepsilon_{max}$ is defined in \eqref{eq: notations}.
\end{proposition}

\noindent\textbf{Proof}. For any $\delta>0$,  from the definition of quantile function for distribution $G_\delta$,  obviously,
 $G_\delta \in\mathcal{M}(\mu,\sigma)$. Next, we show that $d_W(F,G_{\delta})$ is uniformly bounded with respect to $\delta$.
 
 Indeed, for any given $\delta>0$, 
\begin{align*}
 {d^2_W(F,G_{\delta})}&=\int_0^1 {(F^{-1}(u)-G_{\delta}^{-1}(u))}^2\text{d}u\\
		&=(\mu_F-\mu)^2+(\sigma_F-\sigma)^2+2\sigma\sigma_F(1-\text{corr}(F^{-1}(U),G_{\delta}^{-1}(U)))\\
&=\varepsilon_{min}+2\sigma\sigma_F\Big(1-\text{corr}\big(F^{-1}(U),\gamma(U)+2{\delta }F^{-1} (U)\big)\Big).
\end{align*} 
For the term of correlation function, for all $\delta>0$, define 
\begin{align}\label{eq:fdelta}
f(\delta):=\text{corr}(F^{-1}(U),\gamma(U)+2{\delta }F^{-1} (U))
=\frac{2\delta\sigma_F^2+\sigma_0\sigma_F\rho}{\sigma_F\sqrt{\sigma_0^2+4\delta^2\sigma_F^2+4\delta\sigma_0\sigma_F\rho}}.
    \end{align}
One can verify that $f(\delta)$ is (strictly, when $\rho<1$) increasing with respect to $\delta$ on $(0,+\infty)$. Then, we can find that 
$$\lim_{\delta\rightarrow+\infty}f(\delta)=1 \text{~~and~~}\lim_{\delta\rightarrow 0}f(\delta)=\rho.$$
Therefore, we get the upper bound and lower bound, respectively. \hfill$\Box$



\subsection{$\mathcal{N}=\mathcal{M}_{\varepsilon}(\mu,\sigma)$}\label{subsec:3.2}

The distance between the required distribution and the known reference distribution should not be too large. Therefore, this subsection will consider $\mathcal{N}=\mathcal{M}_{\varepsilon}(\mu,\sigma)$ with a given penalty function.  

Firstly, applying the results of Proposition \ref{pro:emax} and Theorem \ref{t1}, we can easily get the following theorem.  





\begin{theorem}\label{tu}
    Let $g$ be a concave distortion function and $\mathcal{N}=\mathcal{M}_{\varepsilon}(\mu,\sigma)$. Suppose  \ref{a1} and \ref{a2}  hold.  For the problem \eqref{m1}, then we have the following situations.
\begin{itemize}	
\item[(i)] Case of $\varepsilon<\varepsilon_{min}$.  In this case, $\mathcal{M}_{\varepsilon}(\mu,\sigma)=\emptyset$ and the problem is meaningless.
\item[(ii)] Case of $\varepsilon= \varepsilon_{min}$. Then  $\mathcal{M}_{\varepsilon}(\mu,\sigma)$ contains only one element, and the optimal quantile expression of $G$ is
\begin{equation*}
	G^{-1}(u)=\mu-\dfrac{\sigma}{\sigma_F}\mu_F +\dfrac{\sigma}{\sigma_F}F^{-1} (u), ~~0<u<1.
\end{equation*}
The corresponding optimal value function is
\begin{equation*}
	H_g (G)-{\varphi({d^2_W(F,G)} )}=\mu+\sigma\sigma_0\rho-{\delta }\varepsilon_{min}.
\end{equation*}
\item[(iii)] Case of $\varepsilon\geq\varepsilon_{max}$.  The problem reduces to Theorem \ref{t1}. 
\end{itemize}
\end{theorem}

\noindent\textbf{Proof}.  By \eqref{eq:emin-emax}, we know that the results of case (i) is obvious. 

For the case (ii), 
when $\varepsilon = \varepsilon_{min}$, then $\mathcal{M}_{\varepsilon}(\mu ,\sigma)$ contains only one element $G$ with $\text{corr}(F^{-1}(U),G^{-1}(U))=1$.  Taking account of the moment constraints, $G$ should have the quantile expression as follows:
\begin{equation*}
	G^{-1}(u)=\mu-\dfrac{\sigma}{\sigma_F}\mu_F+\dfrac{\sigma}{\sigma_F}F^{-1} (u),~~0<u<1.
\end{equation*}
The optimal value is
\begin{align*}
		H_g (G)-{\varphi({d_W^2(F,G)} )}&=\mathbb{E}[\gamma(U)\cdot G^{-1}(U)]-\delta\varepsilon_{min}\\
		&=\mu+\frac{\sigma}{\sigma_F}(  \int_0^1 \gamma(u)F^{-1} (u){\rm d}u-\mu_F )-{\delta }\varepsilon_{min}\\
   &=\mu+\sigma\sigma_0\rho-{\delta }\varepsilon_{min}.
\end{align*}

For the case (iii), by Proposition 
\ref{pro:emax}, we find that the optimal solution $G$, defined by \eqref{eq:optimal-quantile} in Theorem \ref{t1}, lies in  $\mathcal{M}_\varepsilon(\mu,\sigma)$ since $\varepsilon\geq\varepsilon_{max}$.
Hence, when $\varepsilon\geq\varepsilon_{max}$, the following two problems are equivalent: $$\mathop{\sup}_{G\in \mathcal{M}_{\varepsilon}(\mu,\sigma)}\ H_g (G)-\delta {d_W^2(F,G)} = \mathop{\sup}_{G\in \mathcal{M}(\mu,\sigma)}\ H_g (G)-\delta d_W^2(F,G).$$Therefore, 
the problem reduces to Theorem \ref{t1}.  The proof is complete.
\hfill$\Box$

The following technical lemma provides the existence of a distribution with a specific structure in $\mathcal{M}_\varepsilon(\mu,\sigma)$ for the given Wasserstein distance. It is derived from the further discussion and analysis of Proposition \ref{pro:emax}, which will play a key role for Theorem \ref{tu-2}.

\begin{lemma}\label{l1}  Let $g$ be a concave distortion function. 
Suppose \ref{a1} and \ref{a2} hold and $\rho<1$.  For any $\varepsilon\in(\varepsilon_{min},\varepsilon_{max})$, then there exists a unique $\delta^*=\delta^*(\varepsilon)>0$ 
and $G_{\delta^*}\in \mathcal{M}_{\varepsilon}(\mu,\sigma)$ with $d^2_W(F, G_{\delta^*})=\varepsilon$, and its quantile function in the following expression
$$G^{-1}_{\delta^*}(u):=\mu-\dfrac{\sigma}{\sigma_{\delta^*}}\mu_{\delta^*}+\dfrac{\sigma}{\sigma_{\delta^*}}(\gamma(u)+2{{\delta^*} }F^{-1} (u)),~~u\in(0,1),$$
where $\mu_{\delta^*}=\mathbb{E}[\gamma(U)+2\delta^* F^{-1} (U)]$, $\sigma _{\delta^*}=\text{std}(\gamma(U)+2\delta^* F^{-1} (U))$.
To be more specifically,  
\begin{equation}\label{eq:delta*}
\delta^*=-\frac{\sigma_0\rho}{2\sigma_F}+\frac{\sigma_0(\varepsilon_{min}+2\sigma\sigma_F-\varepsilon)\sqrt{1-\rho^2}}{2\sigma_F\sqrt{(\varepsilon_{min}+4\sigma\sigma_F-\varepsilon)(\varepsilon-\varepsilon_{min})}}.
\end{equation}
\end{lemma}

\noindent\textbf{Proof}.
 For any $\varepsilon\in(\varepsilon_{min},\varepsilon_{max})$, for $\delta>0$ and $G_\delta$ defined by \eqref{eq:optimal-quantile-g-delta}, we will choose $\delta^*$ by
setting 
$$d^2_W(F, G_{\delta^*})=\varepsilon.$$

By
virtue of the definition of function $f(\cdot)$ by \eqref{eq:fdelta} in Proposition \ref{pro:emax} and the fact that $f(\delta)$ is strictly increasing with respect to $\delta$ on $(0,+\infty)$. Therefore, there exists a unique $\delta>0$ such that 
\begin{align}\label{eq:fdelta-solution}
f(\delta)=1-\frac{\varepsilon-\varepsilon_{min}}{2\sigma\sigma_F}=\frac{2\delta\sigma_F^2+\sigma_0\sigma_F\rho}{\sigma_F\sqrt{\sigma_0^2+4\delta^2\sigma_F^2+4\delta\sigma_0\sigma_F\rho}}.
\end{align}
By direct calculations, one can solve  $\varepsilon+2\sigma\sigma_F f(\delta)=\varepsilon_{min}+2\sigma\sigma_F$
 and get 
$$\delta^*=\delta^*(\varepsilon)=-\frac{\sigma_0\rho}{2\sigma_F}+\frac{\sigma_0\sqrt{1-\rho^2}(\varepsilon_{min}+2\sigma\sigma_F-\varepsilon)}{2\sigma_F\sqrt{(\varepsilon_{min}+4\sigma\sigma_F-\varepsilon)(\varepsilon-\varepsilon_{min})}}.$$It completes the proof. 
\hfill$\Box$




\begin{remark}\label{rem:3.2}
When $\rho=1$, Lemma \ref{l1} does not hold any more. In this situation, noting that $f(\cdot)$
is defined in \eqref{eq:fdelta}, one can derive that   
$d^2_W(F, G_{\delta})=\varepsilon_{min}$ for all $\delta\geq0$. For $\rho=1$, then the solution of  problem \eqref{m1} is trivial (the same to case (ii) of Theorem \ref{tu}) even for $\varepsilon>\varepsilon_{min}$. Thus, we only consider the situation of $\rho<1$ in the following Theorem \ref{tu-2}. 
\end{remark}

Now we will give the main theorem of this subsection.

\begin{theorem}\label{tu-2}
    Let $g$ be a concave distortion function, and $\mathcal{N}=\mathcal{M}_{\varepsilon}(\mu,\sigma)$, and let \ref{a1} and \ref{a2} hold and $\rho<1$.  Suppose $\varepsilon\in(\varepsilon_{min},\varepsilon_{max})$, and 
$\delta^*>0$ determined by \eqref{eq:delta*}. Then for the problem \eqref{m1}, we have that
    \begin{itemize}
    \item [(i)] When $\delta\ge\delta^*$, problem  \eqref{m1} has a  unique optimal solution and its quantile function is
\begin{equation}\label{eq:hb-th3.3}
G^{-1}_\delta(u)=\mu-\dfrac{\sigma}{\sigma_\delta}\mu_\delta+\dfrac{\sigma}{\sigma_\delta}(\gamma(u)+2{\delta }F^{-1} (u)),~~~0<u<1,
\end{equation}where $\mu_\delta=\mathbb{E}[\gamma(U)+2\delta F^{-1} (U)]$, $\sigma_\delta=\text{std}(\gamma(U)+2\delta F^{-1} (U))$. 
Its corresponding optimal value is
\begin{align*}
H_g (G_\delta)-{\varphi({d_W^2(F,G_\delta)} )}=\mu-\delta(\varepsilon_{min}+2\sigma\sigma_F)+\sigma\sigma_{\delta}.
\end{align*}

\item[(ii)] When $\delta<\delta^*$, problem  \eqref{m1} has a  unique optimal solution and its quantile function is
\begin{equation}
G^{-1}_{\delta^*}(u)=\mu-\dfrac{\sigma}{\sigma_{\delta^*}}\mu_{\delta^*}+\dfrac{\sigma}{\sigma_{\delta^*}}(\gamma(u)+2{\delta^*}F^{-1} (u)),~~~0<u<1.
\end{equation}
  Its corresponding optimal value is
\begin{align*}
		H_g (G_{\delta^*})-\varphi(d_W^2(F,G_{\delta^*}))=\mu-\delta(\varepsilon_{min}+2\sigma\sigma_F)+\sigma\sigma_\delta\text{corr}(\gamma+2\delta F^{-1},G^{-1}_{\delta^*}),
\end{align*}
where $\text{corr}(\gamma+2\delta F^{-1},G^{-1}_{{\delta^*}})=\frac{\sigma_0^2+2\delta\sigma_0\sigma_F\rho+2{\delta^*}\sigma_F\sigma_0\rho+4\delta{\delta^*}\sigma_F\sigma_F}{\sigma_\delta\sigma_{\delta^*}}$.
  
    \end{itemize}

\end{theorem}

\noindent\textbf{Proof}.
 For each $\varepsilon\in(\varepsilon_{min},\varepsilon_{max})$, by Lemma \ref{l1}, we can find
$\delta^*=\delta^*(\varepsilon)$ determined by \eqref{eq:delta*} with $d_W(F,G_{\delta^*})=\varepsilon$. 
Moreover, one can derive that  
\begin{align}\label{eq:delta daoshu}\frac{\mathrm{d}\delta^*(\varepsilon)}{\mathrm{d}\varepsilon}=\frac{\sigma_0\sqrt{1-\rho^2}\Big(-(\varepsilon_{min}+4\sigma\sigma_F-\varepsilon)(\varepsilon-\varepsilon_{min})-(\varepsilon_{min}+2\sigma\sigma_F-\varepsilon)^2\Big)}{2\sigma_F[(\varepsilon_{min}+4\sigma\sigma_F-\varepsilon)(\varepsilon-\varepsilon_{min})]^{\frac{3}{2}}}<0
\end{align}on $(\varepsilon_{min},\varepsilon_{max})$.  It implies that $\delta^*(\cdot)$ is strictly decreasing, then 
 when the penalty function coefficient $\delta$ is fixed, there exists a similarly optimal distance $\varepsilon^*=\varepsilon^*(\delta)$.  


Case (i):   For any 
$\delta\geq \delta^*$, choosing $G^{-1}_\delta$ by \eqref{eq:hb-th3.3}, it implies that 
\begin{align}\label{eq:dw<d}
d_W^2(F, G_\delta)=\varepsilon^*\leq\varepsilon=d_W^2(F, G_{\delta^*}).
\end{align}From Theorem \ref{t1}, we know that 
$$G_\delta=\text{arg}\mathop{\sup}_{G\in \mathcal{M}(\mu,\sigma)}\ H_g (G)-\varphi({d_W^2(F,G)}).$$
By virtue of the fact that $d_W^2(F, G_\delta)=\varepsilon^*\leq\varepsilon$, then we have that 
$$G_\delta=\text{arg}\mathop{\sup}_{G\in \mathcal{M}_{\varepsilon}(\mu,\sigma)}\ H_g (G)-\varphi({d_W^2(F,G)}).$$ Therefore, this case is exactly the same as Theorem \ref{t1}.  

Case (ii): When $\delta<\delta^*$,  from the proof procedure and analysis in case (i),  we know that in this case the optimal distribution  can not be $G_\delta$ in \eqref{eq:hb-th3.3} any more, since 
$$d_W^2(F,G_\delta)=\varepsilon^*(\delta)>\varepsilon=d_W^2(F,G_{\delta^*}),$$resulting in  $G_\delta\notin\mathcal{M}_{\varepsilon}(\mu,\sigma)$.

When $\delta<\delta^*$, we claim that 
\begin{align}\label{eq:g* for delta<delta*}
G_{\delta^*}=\text{arg}\mathop{\sup}_{G\in \mathcal{M}_{\varepsilon}(\mu,\sigma)}\ H_g (G)-\delta {d_W^2(F,G)}.
\end{align}

Step 1: We will show that for any given distance less than $\varepsilon$, the form of \eqref{eq:hb-th3.3} is optimal. 

Indeed, for any given $\bar\varepsilon\leq\varepsilon$ and $\bar\varepsilon\in(\varepsilon_{min},\varepsilon_{max})$, then by Lemma \ref{l1}, there exists a unique $\bar{\delta}^*=\delta^*(\bar{\varepsilon})\geq\delta^*=\delta^*(\varepsilon)$ and $d_W^2(F, G_{\bar{\delta}^*})=\bar\varepsilon$. On the other hand, for any $G\in\mathcal{M}_\varepsilon (\mu,\sigma)$ such that  $d_W(F,G)=\sqrt{\overline{\varepsilon}}$, we obviously have that 
\begin{equation*}	d_W(F,G)=d_W(F,G_{\bar{\delta}^*})\quad\Leftrightarrow\quad \mathbb{E}[F^{-1}\cdot G^{-1}] = \mathbb{E}[F^{-1}\cdot G^{-1}_{\bar{\delta}^*}].
\end{equation*}
From the Cauchy-Schwarz inequality, it implies that $$\mathbb{E}[\gamma\cdot G^{-1}]\le \mathbb{E}[\gamma\cdot G^{-1}_{\bar{\delta}^*}].$$
Therefore,  $G_{\bar{\delta}^*}$ makes the objective function greater than $G$ for the same distance, that is, for any $G\in\mathcal{M}_\varepsilon (\mu,\sigma)$ with  $d_W(F,G)=\sqrt{\overline{\varepsilon}}$, then one has
\begin{align*}
   H_g (G)-\delta d_W^2(F,G)\le H_g (G_{\bar{\delta}^*})-{\delta {d_W^2(F,G_{\bar{\delta}^*})} }. 
\end{align*}
In other words, for any $\bar\varepsilon\leq\varepsilon$ and $\bar\varepsilon\in(\varepsilon_{min},\varepsilon_{max})$, we have that
$$G_{\bar{\delta}^*}=\text{arg}\sup_{G\in\mathcal{M}_\varepsilon (\mu,\sigma)\atop d_W(F,G)=\sqrt{\overline\varepsilon}}H_g (G)-\delta{d_W^2(F,G)}. $$

Step 2: We show that for any $\bar\varepsilon\leq\varepsilon$ and $\bar\varepsilon\in(\varepsilon_{min},\varepsilon_{max})$, $G_{\delta^*}$ is more optimized than $G_{\bar{\delta}^*}$, i.e., \eqref{eq:g* for delta<delta*} holds. Here, $\bar{\delta}^*=\delta^*(\overline\varepsilon)$ and $\delta^*=\delta^*(\varepsilon)$
are determined by \eqref{eq:delta*}, respectively. 

By the one-to-one  corresponding relationship between $\overline\varepsilon$ and $\bar\delta^*$, then showing \eqref{eq:g* for delta<delta*} is equivalent to verify the following fact
 \begin{align}\label{eq:Gstar314-1}
G_{\delta^*}=\text{arg}\mathop{\sup}_{G_{\bar{\delta}^*}\in \mathcal{M}_\varepsilon(\mu,\sigma)\atop \bar{\delta}^*\ge\delta^*>\delta}\quad H_g (G_{\bar{\delta}^*})-\delta{d_W^2(F,G_{\bar{\delta}^*})}.
  \end{align}

Specifically,  for any $\bar{\delta}^*\ge \delta^*> \delta$, we can transform the form of the objective function
\begin{align}\label{Hdelta}
   H_g (G_{\bar{\delta}^*})-\delta d_W^2(F,G_{\bar{\delta}^*})=\mu-\delta(\varepsilon_{min}+2\sigma\sigma_F)+h(\bar{\delta}^*), 
\end{align}where 
\begin{align}\label{hdelta}
h(\bar{\delta}^*)&:=\text{cov}(\gamma+2\delta F^{-1},G^{-1}_{\bar{\delta}^*})\nonumber\\
&=cov(\gamma+2\bar{\delta}^* F^{-1},G^{-1}_{\bar{\delta}^*})+2(\delta-\bar{\delta}^*)\text{cov}(F^{-1},G^{-1}_{\bar{\delta}^*})\nonumber\\
&=\sigma\sigma_{\bar{\delta}^*}+2\frac{(\delta-\bar{\delta}^*)\sigma}{\sigma_{\bar{\delta}^*}}(2\bar{\delta}^*\sigma_F^2+\sigma_0\sigma_F\rho).
\end{align}Here $\sigma_{\bar{\delta}^*}=\sqrt{\sigma_0^2+4(\bar{\delta}^*)^2\sigma_F^2+4\bar{\delta}^*\sigma_0\sigma_F\rho}$. 
The last equality is from the fact that  $\text{corr}(\gamma+2\bar{\delta}^* F^{-1},G^{-1}_{\bar{\delta}^*})=1$ and the formulation of $G^{-1}_{\bar{\delta}^*}$.



By directly calculations, one can derive that
\begin{align*}   \frac{\mathrm{d}h(\bar{\delta}^*)}{\mathrm{d}\bar{\delta}^*}=4\sigma(\delta-\bar{\delta}^*)\frac{\sigma_0^2\sigma_F^2(1-\rho^2)}{\sigma_{\bar{\delta}^*}^3}<0.
\end{align*}
It implies that the objective function is monotonically decreasing with respect to $\bar{\delta}^*$. 

Hence, for any $ \bar{\delta}^*\ge\delta^*>\delta$ and
$G_{\bar{\delta}^*}\in \mathcal{M}_\varepsilon(\mu,\sigma)$, we have that 
$$ H_g (G_{\bar{\delta}^*})-{\delta d_W^2(F,G_{\bar{\delta}^*})} \le H_g (G_{\delta^*})-\delta{d_W^2(F,G_{\delta^*})}.$$In other words,  \eqref{eq:Gstar314-1} holds. 

Furthermore, we can obtain the optimal value of the objective function at $G_{\delta^*}$. 
\begin{align*}
		&~~~H_g (G_{\delta^*})-{\delta({d_W^2(F,G_{\delta^*})} )}\\
&=\mu-\delta(\varepsilon_{min}+2\sigma\sigma_F)+\text{cov}(\gamma+2\delta F^{-1},G^{-1}_{{\delta^*}})\\
&=\mu-\delta(\varepsilon_{min}+2\sigma\sigma_F)+\frac{\sigma}{\sigma_{\delta^*}}\text{cov}(\gamma+2\delta F^{-1},\gamma+2{\delta^*} F^{-1})\\
&=\mu-\delta(\varepsilon_{min}+2\sigma\sigma_F)+\frac{\sigma}{\sigma_{\delta^*}}(\sigma_0^2+2(\delta+\delta^*)\sigma_0\sigma_F\rho+4\delta{\delta^*}\sigma_F^2).
\end{align*}By carefully verifications, we can also get 
$$\text{corr}(\gamma+2\delta F^{-1},G^{-1}_{{\delta^*}})=\frac{\sigma_0^2+2(\delta+\delta^*)\sigma_0\sigma_F\rho+4\delta{\delta^*}\sigma_F^2}{\sigma_\delta\sigma_{\delta^*}}<1.$$ 
The proof is complete. \hfill$\Box$

The results of Theorem \ref{tu-2} are attractive and fascinating. It indicates that the agent makes a \textit{trade-off} between the distortion risk measure and the penalty term, depending on the Wasserstein distance $\varepsilon$ and the penalty parameter $\delta$. This feature is the key ingredient between our model and the model of \cite{bpv24}.  

More precisely, when the penalty parameter $\delta$ is large enough with $\delta>\delta^*$ or the penalty term is dominant, 
then the agent chooses the optimal distribution $G_{\delta}$ such that $$d_W^2(F,G_\delta)<\varepsilon~~~~\text{and}~~~\text{corr}(\gamma+2\delta F^{-1}, G^{-1}_{\delta})=1.$$ In contrast, when the penalty parameter $\delta$  is strictly less than $\delta^*$, then the agent selects the optimal distribution $G_{\delta^*}$ such that $$d_W^2(F,G_{\delta^*})=\varepsilon ~~~~\text{and}~~~ \text{corr}(\gamma+2\delta F^{-1}, G^{-1}_{\delta^*})<1.$$ 

\begin{remark}
When $\delta=0$, then the result of Theorem \ref{tu-2} reduces to Theorem 3.1 in \cite{bpv24}. Theorem \ref{tu-2} shows that the relationship between the penalty parameter and the Wasserstein distance plays a key role in determining the optimal (quantile) distribution. 

Our analysis reveals that the optimal control strategy proposed in \cite{bpv24} ceases to be admissible when the penalty parameter $\delta$ falls below a critical threshold $\delta^*$, particularly when the penalty term is incorporated into the objective functional. This result is exciting, and it also illustrates the non-trivial nature of the study of results with penalty term.
\end{remark}

\begin{figure}[ht]\label{fig:1}
  \centering
  \begin{minipage}{0.45\textwidth}
    \centering
\includegraphics[width=\linewidth]{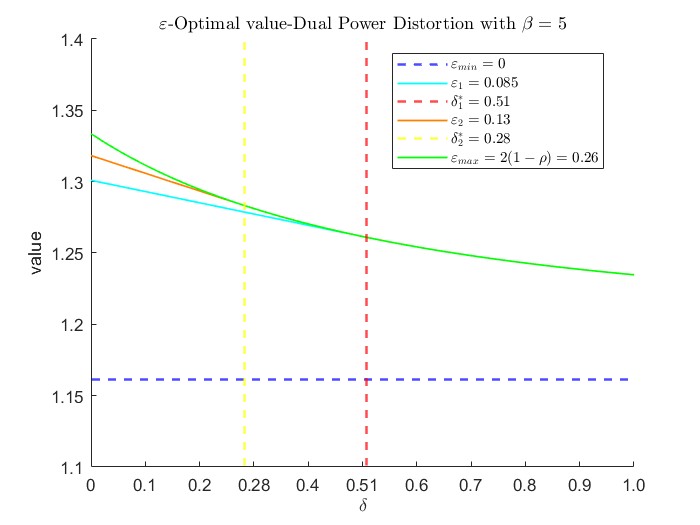}
    \caption*{(a) The optimal value of $\delta$}
    \label{fig:image1}
  \end{minipage}\hfill
  \begin{minipage}{0.45\textwidth}
    \centering
    \includegraphics[width=\linewidth]{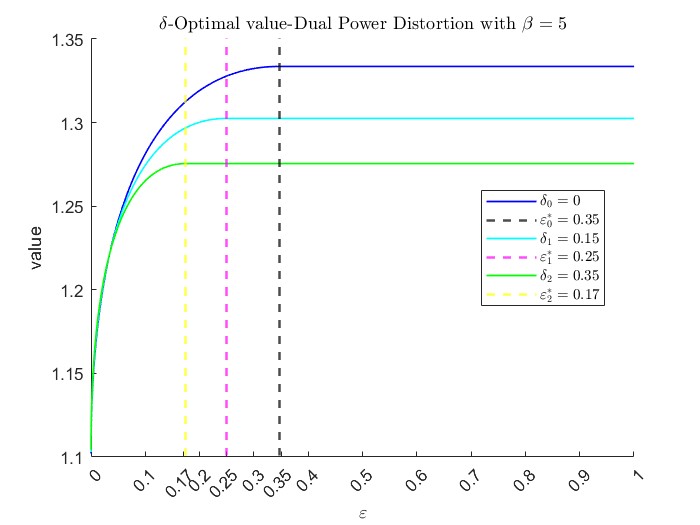}
    \caption*{(b) The optimal value of $\varepsilon$}
    \label{fig:image2}
  \end{minipage}
  \caption*{Figure 1: Dual Power Distortion with $\beta=5$ - Optimal values for different $\delta$ and $\varepsilon$}
  \label{fig:images}
\end{figure}

We end this subsection with some graphic discussion. In order to compare with the results of \cite{bpv24}, 
we choose the reference distribution $F$ as the standard normal distribution with $\mu=\mu_F=0$ and $\sigma= \sigma_F=1$, and the dual power distortion function $g(x)=1-(1-x)^\beta$ with parameter $ \beta = 5 $.

In Figure 1(a), we examine how the optimal value changes with respect to the penalty parameter $\delta$ for different values of $\varepsilon$. The light blue ($\varepsilon_1 = 0.085$) and orange curves ($\varepsilon_2 = 0.13$) correspond to two different distances, with the corresponding critical threshold parameters  $\delta_1^* = 0.51$ (red dashed line) and $\delta_2^* = 0.28$ (yellow dashed line), respectively. The green curve represents the case when the distance reaches the maximum value $\varepsilon_{max} = 0.26$, which reflects the result of Theorem \ref{tu} (iii). It is clear that as the penalty coefficient $\delta$ increases, the resulting optimal value gradually decreases.  This indicates that as the decision-maker's risk aversion (i.e., the penalty coefficient) increases, they select a higher optimal penalty coefficient, leading to a more conservative result. However, when the penalty coefficient becomes too large, the optimal distance gradually converges to the lower bound, consistent with the result of Theorem \ref{tu} (ii). Thus, as the penalty coefficient continues to increase, the optimal value stabilizes and eventually approaches $\varepsilon_{min}$ (blue dashed line), which represents the minimum tolerable distance.

In Figure 1(b), we examine how the optimal value changes with respect to $\varepsilon$ for different values of $\delta$. The blue curve represents the case when $\delta = 0$, where the corresponding optimal distance is $\varepsilon_0 = 0.35$ (black dashed line), which is consistent with the results from \cite{l18}. The light blue and green curves correspond to two different  values of $\delta$, $\delta_1 = 0.15$ (light blue) and $\delta_2 = 0.35$ (green), with corresponding optimal distances of $\varepsilon_1^* = 0.25$ (pink dashed line) and $\varepsilon_2^* = 0.17$ (yellow dashed line), respectively. From the figure, it can be observed that as the value of the penalty function increases, the optimal distance $\varepsilon$ decreases. This phenomenon suggests that as the decision-maker's risk aversion (i.e., the penalty function) becomes stronger, their tolerance for deviations in the distribution decreases, meaning they become more sensitive to deviations in risk.

\begin{figure}[ht]  
  \centering  
  \includegraphics[width=0.5\textwidth]{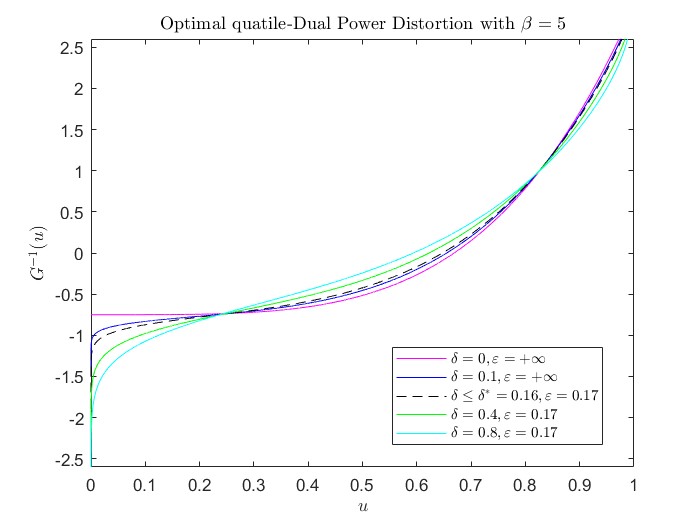}  
  \caption*{Figure 2: Dual Power Distortion with $\beta=5$ - Optimal quantile}  
  \label{fig:image}  
\end{figure}

In Figure 2, we examine the impact of the presence or absence of a penalty term on the optimal quantile when there are no restrictions on the distance. The pink curve represents the results from \cite{l18}, and the blue curve represents the results from Theorem \ref{t1}. Further, we investigate the effect of different penalty coefficients $\delta$ on the optimal quantile when there are distance restrictions. The black dashed line corresponds to Theorem \ref{tu-2} (ii), and encompasses the first case of Theorem 3.1 of \cite{bpv24}, while the light blue and green curves correspond to  Theorem \ref{tu-2} (i).
In this figure, we analyze how  the penalty term affects the selection of the optimal quantile. Specifically, when there are no distance restrictions, the optimal quantile is minimally influenced by the penalty term. However, when distance restrictions are introduced, the size of the penalty coefficient $\delta$ directly impacts the optimal quantile. Through the application of different theorems, we can observe how the optimal quantile changes under varying penalty coefficient conditions.

\section{The general case}\label{sec:4}

In the previous section,  we only consider the cases where the distortion function is concave.
Motivated by \cite{bpv24},  this section will consider the general distortion function $g$, which is no longer a concave function. The isotonic projection technique also works well when the penalty function appears in the objective function. The readers can also refer to \cite{pw24}, where an envelope method is used to solve the non-concave case.

\subsection{General distortion function case: $\mathcal{N}=\mathcal{M}(\mu,\sigma)$}

Define the space of square-integrable, non-decreasing, and left-continuous functions on $(0,1)$ as follows: 
\begin{equation*}
	\mathcal{R}=\left\lbrace {r:(0,1)\to \mathbb{R}~\big|~\int_0^1 r^2(u) \text{d}u<+\infty, ~r\text{~non-decreasing and left-continuous}}\right\rbrace.
\end{equation*}

When $g$ is not concave, then the term $\gamma+2\delta F^{-1}$ is not necessarily increasing, and thereby $G^{-1}$ defined in \eqref{eq:optimal-quantile} maybe not a quantile function any more.   The following definition of isotonic projection can solve this problem.  The readers can refer to Appendix A in \cite{bpv24} for more details and properties of isotonic projections. 

\begin{definition}(Isotonic projection) For $\delta\geq0$, define $\hat{r}_\delta$ as $\gamma+2\delta F^{-1}$ to an isotonic projection on a square-integrable, non-increasing, left-continuous function space on $(0,1)$, i.e
\begin{equation*}
\hat{r}_\delta=\text{arg} \mathop{\min}_{r\in\mathcal{R}}||\gamma+2\delta F^{-1}-r||^2,
\end{equation*}
where $||\cdot||$ denotes the norm on the $L^2$ space.
\end{definition}

\begin{theorem}\label{t41}
Let $g$ be a distortion function and $\mathcal{N}=\mathcal{M}(\mu,\sigma)$. Suppose \ref{a1} and \ref{a2} hold and $\hat{\sigma}_0=\text{std}(\hat{r}_0(U))>0$.  Then problem \eqref{m1} has a unique solution, and its optimal quantile function is
\begin{equation}\label{ep}
	\hat{G}_\delta^{-1}(u)=\mu-\frac{\sigma}{\hat{\sigma}_\delta}\hat{\mu}_\delta+\frac{\sigma}{\hat{\sigma}_\delta}\hat{r}_\delta(u),~~ 0<u<1,
\end{equation}
where $ \hat{\mu}_\delta=\mathbb{E}[\hat{r}_\delta(U)]$. Furthermore, the corresponding optimal value function is
\begin{align*}
H_g (\hat{G}_\delta)-\varphi({d_W^2(F,\hat{G}_\delta)} )=\mu-\delta(\varepsilon_{min}+2\sigma\sigma_F)+\sigma\hat{\sigma}_\delta.
\end{align*}
\end{theorem}

\noindent\textbf{Proof}.  For any $G\in\mathcal{M}(\mu,\sigma)$, by  directly calculations, then the objective function of problem \eqref{m1} can be reduced to 
\begin{align}\label{eg}
H_g (G)-\varphi({d_W^2(F,G)} )
=\mathbb{E}[(\gamma+2\delta F^{-1}(U))\cdot {G}^{-1}(U)]-\delta(\mu_F^2+\mu^2+\sigma_F^2+\sigma^2).
\end{align}

Firstly, it is obvious that $\hat{G}_\delta\in \mathcal{M}(\mu,\sigma)$,  where $\hat{G}_\delta$ is defined by \eqref{ep}. Then by  Lemma B.4 of \cite{bpv24}, we know that 
\begin{align}\label{eq:hatg}
		\mathbb{E}[(\gamma(U)+2\delta F^{-1}(U))\cdot \hat{G}_\delta^{-1}(U)] \ge \mathbb{E}[(\gamma(U)+2\delta F^{-1}(U))\cdot {G}^{-1}(U)], ~~\forall G\in \mathcal{M}(\mu,\sigma).
\end{align}Coming back to \eqref{eg}, one recognizes that $\hat{G}_\delta^{-1}$ is the optimal quantile function.

Noting that $\hat{\sigma}_0>0$, we can easily get  $\hat{\sigma}_\delta>0$ for all $\delta\geq0$.
Furthermore, the corresponding optimal value is
\begin{align*}
		&H_g (\hat{G}_{\delta})-{\varphi({d_W^2(F,\hat{G}_{\delta})} )}\\
		=&\mathbb{E}[\gamma(U)\cdot\hat{G}^{-1}_{\delta}(U)]-\delta\big(\varepsilon_{min}+2\sigma\sigma_F-2\sigma\sigma_F\text{corr}(F^{-1}(U),\hat{G}^{-1}_{\delta}(U))\big)\\
        =&\mu+\sigma \sigma_0\text{corr}(\gamma(U),\hat{r}_\delta(U))-{\delta }(\varepsilon_{min}+2\sigma\sigma_F-2\frac{\sigma}{\hat{\sigma}_{{\delta}}}\text{cov}(F^{-1}(U),\hat{r}_\delta(U))).\\=&\mu+\sigma\sigma_\delta \text{corr}(\gamma(U)+2\delta F^{-1}(U),\hat{r}_\delta(U))-{\delta }(\varepsilon_{min}+2\sigma\sigma_F).
\end{align*}
By the basic properties of the isotonic projection (see Proposition A.3, \cite{bpv24}), we have  $\mathbb{E}[\hat{r}_\delta(U)]=\mathbb{E}[\gamma(U)+2\delta F^{-1}(U)]=1+2\delta\mu_F$ and $\mathbb{E}[(\hat{r}_\delta(U))^2]=\mathbb{E}[(\gamma(U)+2\delta F^{-1}(U))\hat{r}_\delta(U)]$. Hence, it implies that
\begin{align*}
		\text{corr}(\gamma(U)+2\delta F^{-1}(U),\hat{r}_\delta(U))=\frac{\text{cov}(\gamma(U)+2\delta F^{-1}(U),\hat{r}_\delta(U))}{\sigma_\delta\hat{\sigma}_{{\delta}}}
=\frac{\hat{\sigma}_{{\delta}}}{\sigma_\delta}.
\end{align*}
Therefore, the optimal value for $\hat{G}^{-1}_{\delta}$ is
\begin{align*}
		H_g (\hat{G}_{\delta})-{\varphi({d^2_W(F,\hat{G}_{\delta})} )}=\mu+\sigma{\hat{\sigma}_{\delta}}-{\delta }(\varepsilon_{min}+2\sigma\sigma_F).
\end{align*}The proof is complete.  \hfill$\Box$

Theorem \ref{t41} is a general characterization of the optimal quantile under the penalty term situation. It is a natural generalization of Theorem 2 in \cite{l18} and our Theorem \ref{t1}.


Similarly, we introduce the following notations: 
\begin{align}\label{eq: notations-nonconcave}
    \hat{\rho}:=\text{corr}(F^{-1}(U),\hat{\gamma}(U)), 
    \qquad \hat{\sigma}_0=\text{std}(\hat{\gamma}(U)),
 \qquad\hat{\varepsilon}_{max}=\varepsilon_{min}+2\sigma\sigma_F(1-\hat{\rho}).   
\end{align}

In the sequel, we investigate the distance between $\hat{G}_{\delta}$ and $F$.  We add a technical condition for this purpose.

\begin{assumption}\label{a4}
Given the reference distribution $F$ and a distortion function $g$, and suppose $\hat{\sigma}_0>0$ and the following two conditions hold:
\begin{itemize}
    \item[(a)] For any $G\in\mathcal{M}(\mu,\sigma)$,
  $\text{corr}\big(F^{-1}(U),\hat{r}_\delta (U)\big)$ is strictly increasing with $\delta$ on $(0,\infty)$, and
 $$\lim_{\delta\rightarrow 0}\text{corr}\big(F^{-1}(U),\hat{r}_\delta (U)\big)=\hat{\rho} \text{~~and~~} \lim_{\delta\rightarrow+\infty}\text{corr}\big(F^{-1}(U),\hat{r}_\delta (U)\big)=1;$$
    \item[(b)] For any $\delta>0$, $\text{corr}\big(\gamma(U)+\delta F^{-1}(U),\hat{r}_{\delta'} (U)\big)$ is decreasing with respect to $\delta'$ on $(\delta, +\infty)$.
\end{itemize}   
\end{assumption}

\begin{remark}
Suppose $g$ is a concave distortion function and $\rho<1$, then \ref{a4} holds automatically from the proof procedure of Lemma \ref{l1}.  Besides, \ref{a4}(a) also holds in the proof of Theorem 3.7 in \cite{bpv24}.

If $\text{corr}\big(F^{-1}(U),\hat{r}_\delta (U)\big)\equiv 1$ for all $\delta\geq0$, then one can obtain that $d^2_W(F, G_{\delta})=\varepsilon_{min}$ for all $\delta\geq0$. Thus, the solution to problem \eqref{m1} is trivial (the same to Remark \ref{rem:3.2}) even for $\varepsilon>\varepsilon_{min}$. Thus, we only consider the situation of $\hat\rho<1$ in the following Theorem \ref{th-hat}. 
\end{remark}

Following the proof procedure of Proposition \ref{pro:emax} closely, for any distortion function $g$,  suppose \ref{a1}-\ref{a4} hold,  
then one can easily get that 
\begin{align}\label{eq:optimal-distance-general}
\varepsilon_{min}\leq d_W^2(F,\hat{G}_{\delta})\leq \hat{\varepsilon}_{max}, ~~\forall \delta\geq0,
\end{align}where $\hat{\varepsilon}_{max}$ and $\hat{G}_{\delta}$ are defined by \eqref{eq: notations-nonconcave} and \eqref{ep} respectively. Noting that this estimation does not need Assumption  \ref{a4}(b).


\subsection{General distortion function case: $\mathcal{N}=\mathcal{M}_{\varepsilon}(\mu,\sigma)$}

Applying the estimation \eqref{eq:optimal-distance-general}, for the general distortion function $g$ and $\mathcal{N}=\mathcal{M}_{\varepsilon}(\mu,\sigma)$, under Assumptions \ref{a1}-\ref{a4}, then the results of Theorem \ref{tu} also hold similarly. Specifically, when $\varepsilon <\varepsilon_{min}$, then the problem \eqref{m1} has no solution; when $\varepsilon = \varepsilon_{min}$, then the optimal quantile for problem \eqref{m1} is still
\begin{equation*}
	G^{-1}(u)=\mu-\dfrac{\sigma}{\sigma_F}\mu_F+\dfrac{\sigma}{\sigma_F}F^{-1} (u), ~0<u<1,
\end{equation*}
when $\varepsilon\geq\hat{\varepsilon}_{max}$, then the solution to problem \eqref{m1} is the same as in Theorem \ref{t41}.

The following Theorem \ref{th-hat} extends the results of Theorem \ref{tu-2} to the general distortion function and also includes the results of Theorem 3.7 in \cite{bpv24} by incorporating the penalty function.

\begin{theorem}\label{th-hat}
Let $g$ be a distortion function,  $\mathcal{N}=\mathcal{M}_\varepsilon(\mu,\sigma)$ and Assumptions \ref{a1}-\ref{a4} hold.    Suppose $\varepsilon\in(\varepsilon_{min},\hat{\varepsilon}_{max})$.  Then there exists a unique critical threshold $\hat{\delta}^*:=\hat{\delta}^*(\varepsilon)>0$, such that for the problem \eqref{m1}, we claim that 
    
\begin{itemize}
        \item [(i)]When $\delta\ge\hat{\delta}^*$, problem  \eqref{m1} has a  unique optimal solution and its quantile function is
\begin{equation*}
	\hat{G}_\delta^{-1}(u)=\mu-\frac{\sigma}{\hat{\sigma}_\delta}\hat{\mu}_\delta+\frac{\sigma}{\hat{\sigma}_\delta}\hat{r}_\delta(u), ~~0<u<1.
\end{equation*}
Its corresponding optimal value is
\begin{align*}
		H_g (\hat{G}_{\delta})-{\varphi({d_W^2(F,\hat{G}_{\delta})} )}=\mu+\sigma{\hat{\sigma}_{\delta}}-{\delta }(\varepsilon_{min}+2\sigma\sigma_F).
\end{align*}
\item[(ii)] When $\delta<\hat{\delta}^*$, problem  \eqref{m1} has a  unique optimal solution and its quantile function is
\begin{equation}\label{hat*}
\hat{G}_{\hat{\delta}^*}^{-1}(u)=\mu-\dfrac{\sigma}{\hat{\sigma}_{\hat{\delta}^*}}\hat{\mu}_{\hat{\delta}^*}+\dfrac{\sigma}{\hat{\sigma}_{\hat{\delta}^*}}\hat{r}_{\hat{\delta}^*} (u),~~~0<u<1,
\end{equation}where $\hat{\mu}_{\hat{\delta}^*}=\mathbb{E}[\hat{r}_{\hat{\delta}^*}]$ and $\hat{\sigma}_{\hat{\delta}^*}=\text{std}(\hat{r}_{\hat{\delta}^*})$. Its corresponding optimal value is
\begin{align*}
		H_g (\hat{G}_{\hat{\delta}^*})-{\varphi({d^2_W(F,\hat{G}_{\delta^*})} )}=\mu-\delta(\varepsilon_{min}+2\sigma\sigma_F)+\sigma\sigma_\delta\text{corr}(\gamma+2\delta F^{-1},\hat{r}_{\hat{\delta}^*}).
\end{align*}

    \end{itemize}
\end{theorem}

\noindent\textbf{Proof}. 
Similar to Lemma \ref{l1}, for each $\varepsilon\in(\varepsilon_{min}, \hat{\varepsilon}_{max})$, we can find
$\hat{\delta}^*>0$ by setting  $d_W^2(F,\hat{G}_{\delta})=\varepsilon$. From the directly calculation,  then we have  
\begin{align}\label{cdelta}
\text{corr}(F^{-1}(U),\hat{r}_\delta(U))=\frac{\varepsilon_{min}+2\sigma\sigma_F-\varepsilon}{2\sigma\sigma_F}.
\end{align}
By the continuity of the isotonic projection and and Assumption \ref{a4}(a), then there exists a unique $\hat{\delta}^*>0$ such that equation \eqref{cdelta} holds.  From equation \eqref{cdelta} we can see that there is a one-to-one corresponding between $\varepsilon\in(\varepsilon_{min}, \hat{\varepsilon}_{max})$ and $\hat{\delta}^*=\hat{\delta}^*(\varepsilon)$,  and $\hat{\delta}^*$ is inversely proportional to $\varepsilon$.

Conversely, for any $\delta>0$, then there exists a unique $\hat{\varepsilon}^*=\hat{\varepsilon}^*(\delta)$ such that $$\text{corr}(F^{-1}(U),\hat{r}_\delta(U))=\frac{\varepsilon_{min}+2\sigma\sigma_F-\hat{\varepsilon}^*}{2\sigma\sigma_F}.$$

Case (i):  for any 
$\delta\geq \hat{\delta}^*$, choosing $\hat{G}^{-1}_\delta$ by \eqref{ep}, it implies that 
\begin{align}
d_W^2(F, \hat{G}_\delta)=\hat{\varepsilon}^*\leq\varepsilon=d_W^2(F, \hat{G}_{\hat{\delta}^*}).
\end{align}

According to Theorem \ref{t1}, the objective function can also be written as
\begin{align*}
	H_g ({G})+{\varphi({d_W^2(F,G)})}=\mathbb{E}[(\gamma+2\delta F^{-1})\cdot {G}^{-1}]-\delta(\mu_F^2+\mu^2+\sigma_F^2+\sigma^2),
\end{align*}
and  (by Lemma B.4 of \cite{bpv24}) we also have 
\begin{equation*}
	 \mathbb{E}[(\gamma+2\delta F^{-1})\cdot {G}^{-1}] \le \mathbb{E}[(\gamma+2\delta F^{-1})\cdot \hat{G}^{-1}_{\delta}], ~~\forall G \in \mathcal{M}(\mu,\sigma).
\end{equation*}

Therefore,  we know that 
$$\hat{G}_\delta=\text{arg}\mathop{\sup}_{G\in \mathcal{M}(\mu,\sigma)}\ H_g (G)-\varphi({d_W(F,G)}^2).$$
By virtue of the fact that $d_W^2(F, \hat{G}_\delta)=\hat{\varepsilon}^*\leq\varepsilon$, then we have that 
$$\hat{G}_\delta=\text{arg}\mathop{\sup}_{G\in \mathcal{M}_{\varepsilon}(\mu,\sigma)}\ H_g (G)-\delta {d_W^2(F,G)}.$$ 
Moreover, the optimal value of $\hat{G}^{-1}_{\delta}$ is $H_g (\hat{G}_{\delta})-{\varphi({d^2_W(F,\hat{G}_{\delta})})}=\mu+\sigma{\hat{\sigma}_{\delta}}-{\delta }(\varepsilon_{min}+2\sigma\sigma_F)$.

Case (ii): this proof procedure is similar to Theorem \ref{tu-2}.   
When $\delta<{\hat{\delta}^*}$,  from the proof procedure and analysis of case (i),  we know that in this case the optimal distribution  can not be $G_\delta$ in \eqref{eq:hb-th3.3} any more, since 
$
d_W^2(F, \hat{G}_\delta)=\hat{\varepsilon}^*>\varepsilon=d_W^2(F, \hat{G}_{\hat{\delta}^*})$, which means  $\hat{G}_\delta\notin\mathcal{M}_{\varepsilon}(\mu,\sigma)$.

Firstly, we claim that for any $\bar\varepsilon\leq\varepsilon$ and $\bar\varepsilon\in(\varepsilon_{min},\varepsilon_{max})$, we have that
\begin{equation}\label{eq:step1-nonc}
\hat{G}_{\bar\delta}=\text{arg}\sup_{G\in\mathcal{M}_\varepsilon (\mu,\sigma)\atop d_W(F,G)=\sqrt{\overline\varepsilon}}H_g (G)-\delta{d^2_W(F,G)},   
\end{equation}where 
 $\bar{\delta}=\hat{\delta}^*(\bar{\varepsilon})\geq\hat{\delta}^*=\hat{\delta}^*(\varepsilon)$ since $\bar{\varepsilon}\leq \varepsilon$.

In fact, for any given $\bar\varepsilon\leq\varepsilon$ and $\bar\varepsilon\in(\varepsilon_{min},\hat{\varepsilon}_{max})$, since $d_W^2(F, \hat{G}_{\bar\delta})=\bar\varepsilon$, and the fact that for any $G \in \mathcal{M}_{\varepsilon}(\mu,\sigma)$ with $d_W(F,G)=\sqrt{\overline{\varepsilon}}$, we have that 
\begin{equation*}
	 \mathbb{E}[(\gamma+2\delta F^{-1})\cdot {G}^{-1}] \le \mathbb{E}[(\gamma+2\delta F^{-1})\cdot \hat{G}^{-1}_{\bar{\delta}}].
\end{equation*}
Therefore,  $\hat{G}_{\bar{\delta}}$ makes the objective function greater than ${G}^{-1}$ for the same distance, that is, for any $G\in\mathcal{M}_\varepsilon (\mu,\sigma)$ with  $d_W(F,G)=\sqrt{\overline{\varepsilon}}$, then one has
\begin{align*}
   H_g (G)-{\delta({d^2_W(F,G)} )}\le H_g (\hat{G}_{\bar{\delta}})-{\delta {d^2_W(F,\hat{G}_{\bar{\delta}})} }, 
\end{align*}which is exactly \eqref{eq:step1-nonc}. 

Secondly, for any $\bar\varepsilon\leq\varepsilon$ and $\bar\varepsilon\in(\varepsilon_{min},\hat{\varepsilon}_{max})$, we show that 
 \begin{align}\label{eq:Gstar314}
\hat{G}_{\hat{\delta}^*}=\text{arg}\mathop{\sup}_{\hat{G}_{\bar{\delta}}\in \mathcal{M}_\varepsilon(\mu,\sigma)\atop \bar{\delta}\ge\hat{\delta}^*>\delta}\quad H_g (\hat{G}_{\bar{\delta}})-\delta{d_W^2(F,\hat{G}_{\bar{\delta}})},
  \end{align}where 
 $\bar{\delta}=\hat{\delta}^*(\bar{\varepsilon})\geq\hat{\delta}^*=\hat{\delta}^*(\varepsilon)>\delta$ since $\bar{\varepsilon}\leq \varepsilon$.

Specifically,  for any $\delta<\hat{\delta}^*\le\bar{\delta}$, we can transform the form of the objective function as 
\begin{align*}
	H_g (\hat{G}_{\bar{\delta}})-\delta({d_W^2(F,\hat{G}_{\bar{\delta}})})
    =\mu-\delta(\varepsilon_{min}+2\sigma\sigma_F)+h(\bar{\delta}),
\end{align*}
where $h(\bar{\delta})=\text{cov}(\gamma+2\delta F^{-1},\hat{G}^{-1}_{\bar{\delta}})$. 
Moreover, one has 
\begin{align*}
h(\bar{\delta})=\text{cov}(\gamma+2\delta F^{-1},\hat{G}^{-1}_{\bar{\delta}})=\sigma_\delta \sigma \text{corr}(\gamma+2\delta F^{-1},\hat{\gamma}^{-1}_{\bar{\delta}}),
\end{align*} then by the Assumption A3(b), it implies that $h(\bar{\delta})$ is decreasing in $\bar{\delta}$.



Hence, for any $ \bar{\delta}\ge\delta^*>\delta$ and
$\hat{G}_{\bar{\delta}}\in \mathcal{M}_\varepsilon(\mu,\sigma)$, we have that 
$$ H_g (\hat{G}_{\bar{\delta}})-{\delta d_W^2(F,\hat{G}_{\bar{\delta}})} \le H_g (\hat{G}_{\delta^*})-\delta{d_W^2(F,\hat{G}_{\delta^*})}.$$Therefore,  \eqref{eq:Gstar314} holds.  

Finally, combining \eqref{eq:step1-nonc} and \eqref{eq:Gstar314} together, we finally prove that 
\begin{align}\label{hatd*}
\hat{G}_{\hat{\delta}^*}=\text{arg}\mathop{\sup}_{G\in \mathcal{M}_{\varepsilon}(\mu,\sigma)}\ H_g (G)-\delta {d^2_W(F,G)}.
\end{align}

Furthermore, we can obtain the optimal value of the objective function at $\hat{G}_{\delta^*}$. 
\begin{align*}
   & H_g(\hat{G}_{\hat{\delta}^*})-\varphi(d_W^2(F,\hat{G}_{\hat{\delta}^*}))\\
&=\mathbb{E}\big(\gamma\cdot\hat{G}^{-1}_{\hat{\delta}^*}\big)-\varphi[\varepsilon_{min}+2\sigma\sigma_F-2\sigma\sigma_F\text{corr}(F^{-1},\hat{G}^{-1}_{\hat{\delta}^*})]\\
&=\mu+\frac{\sigma}{\hat{\sigma}_{\hat{\delta}^*}}(\int_0^1 \gamma(u)\hat{r}_{\hat{\delta}^*}(u){\rm d}u- \hat{\mu}_{\hat{\delta}^*})-{\delta }(\varepsilon_{min}+2\sigma\sigma_F-2\text{cov}(F^{-1},\hat{G}^{-1}_{\hat{\delta}^*}) )\\
&=\mu+\sigma\sigma_0\text{corr}(\gamma,\hat{r}_{\hat{\delta}^*})-{\delta }(\varepsilon_{min}+2\sigma\sigma_F-2\frac{\sigma}{\hat{\sigma}_{\hat{\delta}^*}}\text{cov}(F^{-1},\hat{r}_{\hat{\delta}^*}))\\
&=\mu+\sigma\sigma_\delta \text{corr}(\gamma+2\delta F^{-1},\hat{r}_{\hat{\delta}^*})-{\delta }(\varepsilon_{min}+2\sigma\sigma_F),
\end{align*}where $\hat{r}_{\delta^*}$is the projection of $\gamma+2{\delta^*}F^{-1}$,  $\hat{\mu}_{\hat{\delta}^*}=\mathbb{E}[\hat{r}_{\delta^*}]$ and $\hat{\sigma}_{\hat{\delta}^*}=\text{std}(\hat{r}_{\delta^*})$. 
\hfill$\Box$

Theorem \ref{th-hat}  shows that when the penalty term is involved, there exist similar characterizations of Theorem \ref{tu-2} under the general distortion function.

The following corollary is almost obvious. The proof is omitted.

\begin{corollary}\label{c41}
($\varepsilon=+\infty$)\quad Suppose \ref{a1}-\ref{a4} hold and $g$ is a distortion function. Then
\begin{align*}
		\mathop{\sup}_{G\in \mathcal{M}(\mu,\sigma)}H_g(G)-\varphi(d_W^2(F,{G}))=\mu+\sigma\sigma_\delta -{\delta }(\varepsilon_{min}+2\sigma\sigma_F ).
\end{align*}
\end{corollary}


\section{An illustrate example of CVaR}\label{sec:5}

In this section, we apply our results to one widely used risk measure, called Conditional Value at Risk (CVaR).  We briefly introduce these as follows. The readers can refer to \cite{a02} and \cite{fs16} for more details.   

For any given $G\in\mathcal{M}^2$, 
 Value at Risk (VaR) is defined as
\begin{equation*}
\text{VaR}_\alpha(G):=G^{-1}(\alpha), ~~\alpha\in(0,1),
\end{equation*}
and its corresponding distortion function is $g(x)=\mathbf{1}_{(1-\alpha,1]}(x), ~~x\in[0,1].$
Conditional Value at Risk (CVaR) (also called Expected Shortfall (ES)) is denoted by
\begin{equation*}
	\text{CVaR}_{\alpha}(G):=\frac{1}{1-\alpha}\int_\alpha^1 \text{VaR}_u(G)\text{d}u =\int_0^1 G^{-1}(u)\text{d}{g}(u),
\end{equation*}
where the corresponding distortion function is $g(x)=\min\left\lbrace  \frac{x}{1-\alpha},1\right\rbrace, ~x\in[0,1],$ 
and the derivative of the distortion function is
$\gamma(x)=\frac{1}{1-\alpha}\mathbf{1}_{[0,1-\alpha]}(x)$.  

The following proposition establishes the optimal quantiles and optimal values for CVaR. 

\begin{proposition}\label{6.1}
For $\alpha\in(0, 1)$, $\varepsilon\in(\varepsilon_{min},\varepsilon_{max})$ and $\delta\geq0$,  we have the optimal penalty coefficient as $\delta^*$, where $\delta^*$ satisfies
\begin{equation*}
\delta^*=-\frac{\rho\sqrt{\frac{\alpha}{1-\alpha}}}{2\sigma_F}+\frac{(\varepsilon_{min}+2\sigma\sigma_F-\varepsilon)\sqrt{\frac{\alpha(1-\rho^2)}{1-\alpha}}}{2\sigma_F\sqrt{(\varepsilon_{min}+4\sigma\sigma_F-\varepsilon)(\varepsilon-\varepsilon_{min})}}.
\end{equation*}
\begin{itemize}
    \item [(i)]When $\delta\ge\delta^*$, the value of CVaR with penalty under distribution uncertainty is
    \begin{align*}
    \mathop{\sup}_{G\in\mathcal{M}_{\varepsilon}(\mu,\sigma)}\text{CVaR}_{\alpha}(G)-\delta d^2_W(F,G)=\mu+\sigma\sigma_{CVaR_\delta} -{\delta }(\varepsilon_{min}+2\sigma\sigma_F ),
    \end{align*}
    and the optimal quantile function  is 
   \small{ $${G}^{-1}_{CVaR_\delta}(u)=\mu+ \left(\frac{\sigma(2\delta F^{-1}(u)-1-2\delta\mu_F)}{\sigma_{{CVaR_\delta}}}\right)\mathbf{1}_{u\in(0,\alpha]}+\left(\frac{\sigma(\frac{\alpha}{1-\alpha}+2\delta F^{-1}(u)-2\delta\mu_F)}{\sigma_{{CVaR_\delta}}}\right) \mathbf{1}_{u\in(\alpha,1)},$$}
    where $\sigma_{CVaR_\delta}=\sqrt{\frac{\alpha}{1-\alpha}+4\delta^2(\mu_F^2+\sigma_F^2) +4\delta \sigma_F\rho\sqrt{\frac{\alpha}{1-\alpha}}.}$
    \item [(ii)]When $\delta<\delta^*$, the value of CVaR with penalty under distribution uncertainty is
    \begin{align*}
    \mathop{\sup}_{G\in\mathcal{M}_{\varepsilon}(\mu,\sigma)}\text{CVaR}_{\alpha}(G)-\delta d^2_W(F,G)=\mu+\sigma\sigma_{CVaR_{\delta^*}} -{\delta }(\varepsilon_{min}+2\sigma\sigma_F ),
    \end{align*}
    and the optimal quantile function  is 
    \small{$${G}^{-1}_{CVaR_{\delta^*}}(u)=\mu+ \left(\frac{\sigma(2\delta^* F^{-1}(u)-1-2\delta^*\mu_F)}{\sigma_{CVaR_{\delta^*}}}\right)\mathbf{1}_{u\in(0,\alpha]}+\left(\frac{\sigma(\frac{\alpha}{1-\alpha}+2{\delta}^* F^{-1}(u)-2{\delta}^*\mu_F)}{\sigma_{CVaR_{\delta}^*}}\right) \mathbf{1}_{u\in(\alpha,1)},$$}
    where $\sigma_{CVaR_{\delta^*}}=\sqrt{\frac{\alpha}{1-\alpha}+4{{\delta}^*}^2(\mu_F^2+\sigma_F^2) +4{\delta}^* \sigma_F\rho\sqrt{\frac{\alpha}{1-\alpha}}.}$
\end{itemize}

\end{proposition}

\noindent\textbf{Proof}.
When $\delta<\delta^*$, since the derivative function of distortion function for CVaR is $\gamma(u)=\frac{1}{1-\alpha}\mathbf{1}_{[0,1-\alpha]}(u)$, $u\in (0,1)$, which is non-decreasing,  applying Theorem \ref{t1}, then we have
\begin{align*}
\mathop{\sup}_{G\in\mathcal{M}(\mu,\sigma)}\text{CVaR}_{\alpha}(G)-\delta d^2_W(F,G)=\mu+\sigma\sigma_{CVaR} -{\delta }(\varepsilon_{min}+2\sigma\sigma_F ), 
\end{align*}
where 
$\sigma_{CVaR_\delta}^2:=var(\gamma(U)+2\delta F^{-1}(U))=\frac{\alpha}{1-\alpha}+4\delta^2(\mu_F^2+\sigma_F^2) +4\delta \sigma_F\rho\sqrt{\frac{\alpha}{1-\alpha}}.$

Then the optimal quantile of CVaR is
\begin{align*}
		{G}^{-1}_{CVaR}=\mu+ [\frac{\sigma(2\delta F^{-1}(u)-1-2\delta\mu_F)}{\sigma_{CVaR}}]\mathbf{1}_{u\in(0,\alpha]}+[\frac{\sigma(\frac{\alpha}{1-\alpha}+2\delta F^{-1}(u)-2\delta\mu_F)}{\sigma_{CVaR}}] \mathbf{1}_{u\in(\alpha,1)}.
\end{align*}

When $\delta<\delta^*$, we apply Lemma \ref{l1}, and we obtain

\begin{equation*}
    \sigma_0^2=\int_0^1 (\gamma(u)-1)^2\text{d}u=\frac{\alpha}{1-\alpha},
\end{equation*}
then we have the optimal penalty coefficient as $\delta^*$ , where $\delta^*$ satisfies

\begin{equation*}
\delta^*=-\frac{\rho\sqrt{\frac{\alpha}{1-\alpha}}}{2\sigma_F}+\frac{(\varepsilon_{min}+2\sigma\sigma_F-\varepsilon)\sqrt{\frac{\alpha(1-\rho^2)}{1-\alpha}}}{2\sigma_F\sqrt{(\varepsilon_{min}+4\sigma\sigma_F-\varepsilon)(\varepsilon-\varepsilon_{min})}}.
\end{equation*}

\hfill$\Box$

Finally, we do some numerical analysis for the situation of CVaR. 
We choose the reference distribution $F$ as the standard normal distribution and let $\mu=\mu_F=0$, $\sigma= \sigma_F=1$ and $\alpha=0.7$. Figure 3 (a) reflects how the optimal value changes with respect to $\delta$ for different values of $\varepsilon$. The light blue and orange curves correspond to two different values of $\varepsilon$, $\varepsilon_1 = 0.16$ (light blue) and $\varepsilon_2 = 0.32$ (orange), with $\delta_1^* = 0.86$ (red dashed line) and $\delta_2^* = 0.40$ (yellow dashed line), respectively. The green curve represents the case when the distance reaches the maximum value $\varepsilon_{\text{max}} = 0.26$.

\begin{figure}[ht]
  \centering
  \begin{minipage}{0.45\textwidth}
    \centering
    \includegraphics[width=\linewidth]{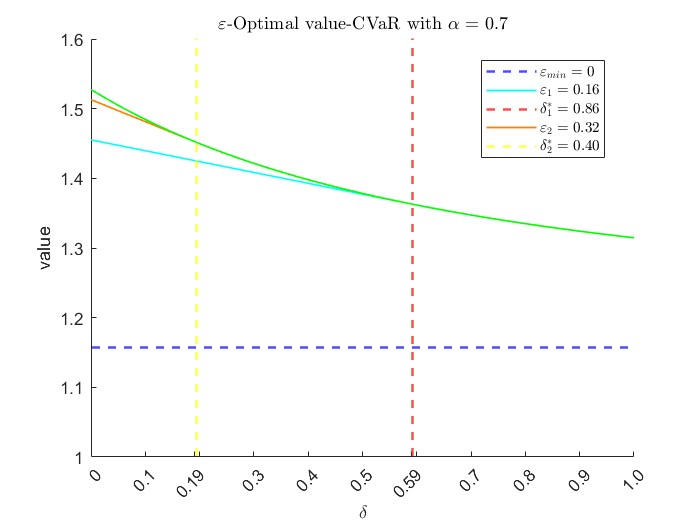}
    \caption*{(a) The optimal value of $\delta$}
    \label{fig:image1}
  \end{minipage}\hfill
  \begin{minipage}{0.45\textwidth}
    \centering
    \includegraphics[width=\linewidth]{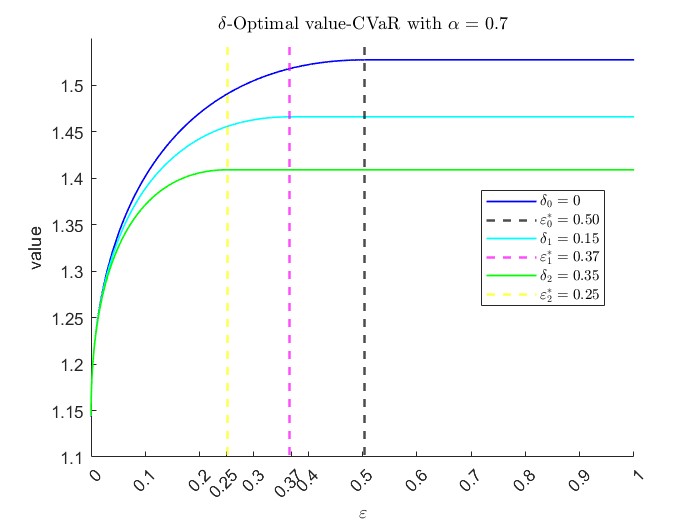}
    \caption*{(b) The optimal value of $\varepsilon$}
    \label{fig:image2}
  \end{minipage}
  \caption*{Figure 3: CVaR with $\alpha=0.7$ - Optimal values for different $\delta$ and $\varepsilon$}
  \label{fig:images}
\end{figure}

Figure 3 (b) displays how the optimal value changes with respect to $\varepsilon$ for different values of $\delta$. The blue curve represents the case when $\delta = 0$, where the corresponding optimal distance is $\varepsilon_0 = 0.50$ (black dashed line). The light blue and green curves correspond to two different values of $\delta$,  $\delta_1 = 0.15$ (light blue) and $\delta_2 = 0.35$ (green), with corresponding optimal distances of $\varepsilon_1^* = 0.37$ (pink dashed line) and $\varepsilon_2^* = 0.25$ (yellow dashed line), respectively.

\begin{figure}[ht]  
  \centering  
  \includegraphics[width=0.5\textwidth]{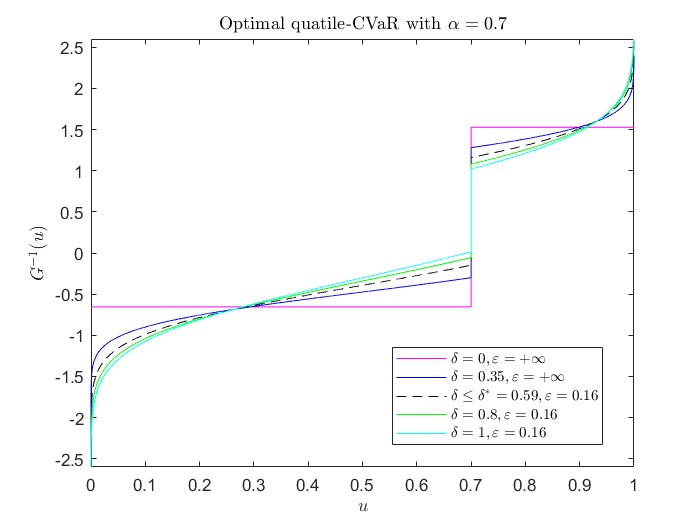}  
  \caption*{Figure 4: CVaR with $\alpha=0.7$ - Optimal quantile}  
  \label{fig:image}  
\end{figure}

Figure 4 shows the impact of the presence or absence of a penalty term on the optimal quantile when there are no restrictions on distance. The pink curve shows the result of \cite{l18}, and the blue curve represents the results from Theorem 3.1. Further, we study the effect of different penalty coefficients $\delta$ on the optimal quantile. The black dashed line corresponds to Theorem \ref{tu-2} (ii), and includes the first case of Theorem 3.1 in \cite{bpv24}, while the light blue and green curves correspond to Theorem \ref{tu-2} (i).

\section{Conclusions}\label{sec:6}
In this paper, we focus on studying the distortion risk measure with a linear penalty function under distributional uncertainty.  This modification allows us to deeply explore the impact of the  Wasserstein distance and the penalty parameter together,  which reveals that  the agent makes a trade-off between the distortion risk measure and the penalty term.   This characteristic is the key ingredient between our model and the models of \cite{bpv24} and \cite{l18}.    Our findings extend  the corresponding results presented in \cite{bpv24} and \cite{l18}, notably adding the need for penalty terms in the framework of robust distortion risk measures.

\section*{Acknowledgments}

The authors thank the grant supported by the Fundamental Research Funds for the Central Universities (No. 2024KYJD2008).

\newpage

\bibliographystyle{jf}

\end{document}